\documentclass[sigconf]{acmart}
\renewcommand\footnotetextcopyrightpermission[1]{} 
\usepackage{graphicx}
\usepackage{amsmath}
\usepackage{booktabs}
\usepackage{amsfonts}
\usepackage{multirow}[c]
\usepackage{makecell}
\usepackage{subfig}
\usepackage{color}
\usepackage{bm}
\usepackage{epstopdf}
\usepackage{url}
\usepackage[cal=cm]{mathalfa}
\usepackage{balance}
\usepackage{threeparttable}
\usepackage{wrapfig}
\usepackage{tabularx}
\usepackage{array}
\usepackage{enumitem}
\usepackage{appendix}
\usepackage{tikz}
\usepackage{float} 
\usepackage{adjustbox}
\usepackage[linesnumbered,ruled,vlined]{algorithm2e}

\SetKwProg{Stage}{Stage}{}{} 

\setlist[itemize]{leftmargin=*}

\AtBeginDocument{}

\title{Break the Inaccessible Boundary: Distilling Post-Conversion Content for User Retention Modeling}

\setcopyright{acmcopyright}
\copyrightyear{2026}
\acmYear{2026}
\acmDOI{XXXXXXX}

\acmConference[Conference acronym 'XX]{Make sure to enter the correct
  conference title from your rights confirmation email}{June 03--05,
  2018}{Woodstock, NY}

\author{Tianbao Ma*, Ruochen Yang*, Chengen Li, Yuexin Shi, Jiangxia Cao$^\dagger$, \\Linxun Chen, Zhaojie Liu, Yanan Niu, Han Li and Kun Gai}
\thanks{* Equal Contributions, Jiangxia Cao is the corresponding author.}
\affiliation{
  \institution{Kuaishou Technology, Beijing, China}
 \country{\{matianbao, yangruochen, lichengen, shiyuexin, caojiangxia, chenxi36, zhaotianxing, niuyanan, lihan08\}@kuaishou.com, gai.kun@qq.com}
}

\begin{document}

\begin{abstract}

User retention is a key metric to measure long-term engagement in modern platforms.
In real-time bidding (RTB) advertising system for user re-engagement, the retention model is required to predict future revisit probability at bidding time, before the user converts and consumes any content.
Although post-conversion content, termed \textbf{\textit{Onboarding Content}}, provides highly informative signals for retention prediction, directly using it in training causes severe feature leakage and creates a gap between training and serving.
To address this issue, we propose \textbf{OCARM}, a two-stage distillation-aligned framework for \textbf{\textit{O}}nboarding \textbf{\textit{C}}ontent \textbf{\textit{A}}ugmented \textbf{\textit{R}}etention \textbf{\textit{M}}odeling,
enabling the model to implicitly capture future content using only observable features during inference. 
In the first stage, we deliberately expose onboarding content to train a hierarchical encoder that produces teacher representations. 
In the second stage, a user encoder is aligned with the frozen teacher through distillation, allowing the model to approximate the inaccessible onboarding signals without leakage. 
Extensive offline experiments and online A/B tests demonstrate that our framework achieves consistent improvements in a real-world growth scenario.

\end{abstract}

\ccsdesc[500]{Information systems~Recommender systems}

\keywords{Recommender systems, Retention Modeling, Feature Leakage}

\maketitle

\section{Introduction}


User retention serves as a critical metric for assessing long-term user engagement, capturing the sustained value of user-platform interactions and representing a core optimization objective in modern recommender systems~\cite{onelive} and user growth scenarios~\cite{pushgen}.
Unlike immediate short-term feedback signals such as clicks and stays, retention more accurately reflects users' overall satisfaction and value perception, indicating whether a platform can consistently deliver sufficient value to drive sustained revisits and realize a positive return on marketing investment.
Consequently, in real-time bidding (RTB) advertising systems designed for user re-engagement, the predictive accuracy of retention prediction directly determines the decisions regarding bidding strategies, audience targeting and budget allocation, ultimately influencing the conversion efficiency and incremental returns of the overall campaign. 

Optimizing retention modeling has emerged as a pivotal challenge in industrial user growth, with increasing research in recent years~\cite{DT4Rec, GFN4Retention, RLUR, peak_end}. 
Despite diverse technical approaches, existing methods largely share the same idea, which is to enhance long-term retention by optimizing recommendation policies to deliver more relevant and engaging content.
These approaches are predominantly deployed in the post-acquisition phase, where users have already entered the platform and their interactions are managed by recommendation systems. 
In this setting, the platform possesses abundant prior user profiles and behavior contexts, leveraging them as the input features for the model to optimize retention targets.

\begin{figure}[t!]
\begin{center}
\includegraphics[width=0.95\linewidth]{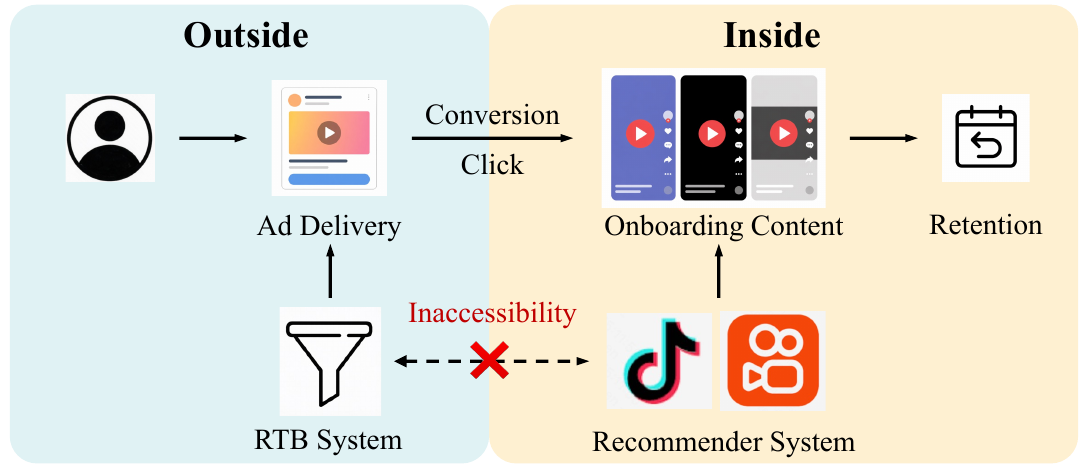} 
\vspace{-0.3cm}
\caption{The journey of user retention, where bidding decision is made prior to conversion, and the onboarding content is inaccessible at this time.}
\vspace{-0.5cm}
\label{fig:motivation}
\end{center}
\end{figure}

However, in the scenarios of user re-engagement driven by advertisements, this optimization process undergoes a fundamental difference due to the inaccessibility of information.
As shown in Figure \ref{fig:motivation}, the retention prediction model included in RTB system operates at the bidding stage.
At this upstream growth decision point, the system is required to forecast a user's subsequent revisit probability and retention trajectory without entering the platform or consuming any content.
The user journey proceeds through ad delivery, conversion (click to open app), and after which the user begins to interact with in-app content.
We define the post-conversion content consumption experience, that is, the videos watched or the interactions performed, as the \textbf{\textit{Onboarding Content}}.
Consequently, a temporal inaccessibility boundary naturally emerges at the conversion node. Information to the left of this boundary, including static user profiles, historical behavior sequences and ad context, is fully observable at decision time.
While to the right, encompassing the onboarding content and its derived interaction feedback, constitutes unobserved future signals that are strictly inaccessible during the bidding stage.

Notably, the onboarding content on the future side of this boundary carries exceptionally powerful predictive signals for retention.
Consistent with prior work that incorporates such content as the vehicle for observational inputs or reward features in retention prediction, we conduct a feature leakage experiment by deliberately injecting onboarding content into training, and observe a substantial AUC gain (Section \ref{subsec:offline}). 
This confirms that the content user experiences after returning exerts a strong predictive influence on their subsequent revisit~\cite{foresight}. 
However, precisely due to the high information density and strong posterior correlation of this signal, directly incorporating it into training induces severe feature leakage.
The model learns to rely on future signals that are systematically absent during inference, resulting in an irreconcilable distribution gap between training and serving.

Consequently, the core challenge lies in bridging the information barrier between growth system and recommendation system, that is, enabling downstream content supply to inversely guide upstream bidding and conversion decisions.
A natural intuition is to feed user features into the recommendation model at bidding time to predict pseudo labels for potential content interactions. 
However, preference mismatch and the prohibitive cost under real-time bidding constraints prevents the acquisition of explicit content.

Nevertheless, we observe that although the exact onboarding content remains unknown at bidding time, user's static profiles and historical preferences implicitly constrain the distribution of content consuming after re-engaging.
This constraint arises since the onboarding content recommendation heavily depend on the same observable priors either.
Therefore, learning a mapping from observable user features into the latent space of onboarding content representations allows the model to approximate feature interactions, without requiring direct access to unobserved information.

To this end, we propose a two-stage distillation-aligned framework \textbf{OCARM} for \textbf{\textit{O}}nboarding \textbf{\textit{C}}ontent \textbf{\textit{A}}ugmented \textbf{\textit{R}}etention \textbf{\textit{M}}odeling, enabling the retention model to implicitly perceive future consumed content at inference time without feature leakage, thus providing high-quality signals for model optimization.
In the first stage, we intentionally allow feature leakage during training by jointly encoding future interaction sequence logs together with the retention model inputs, and train an onboarding content encoder to produce teacher representations. 
The encoder employs hierarchical attention to compress sequential content representations while capturing both intra-day and inter-day temporal causality.
In the second stage, a user encoder is trained to align with the frozen onboarding content representations via distillation objective, thereby approximating unobserved future features.
Therefore, at inference time, the model requires only observable features for informative retention prediction for bidding decisions.

We conduct extensive offline experiments to evaluate the effectiveness of onboarding content distillation in bridging growth system and recommendation system. 
The A/B tests also demonstrate the value of our approach in a real growth scenario.

\section{Methodology}

\begin{figure*}[t!]
\begin{center}
\includegraphics[width=0.9\linewidth]{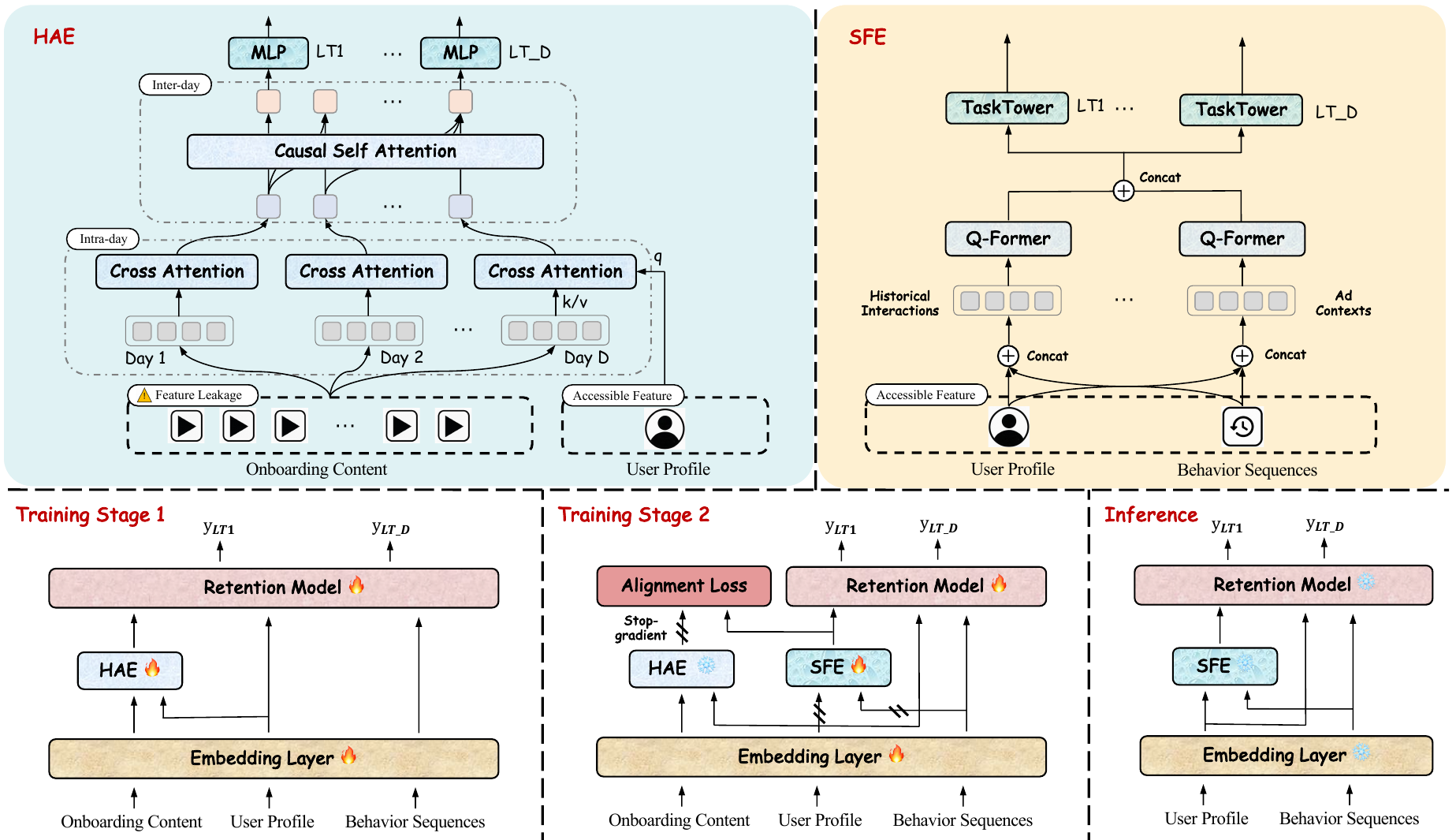} 
\vspace{-0.2cm}
\caption{Overview of the proposed framework OCARM. In Stage 1, HAE learns teacher representations from deliberately leaked onboarding content. In Stage 2, SFE is trained to align with the frozen teacher through distillation based on accessible user-side features. During inference, only SFE is activated for retention prediction without feature leakage.}
\vspace{-0.3cm}
\label{fig:model}
\end{center}
\end{figure*}

As the overview depicted in Figure~\ref{fig:model}, by deliberately leaking the onboarding content and encoding to obtain distilled signals, we align the user representation and enable informative inference. 

\subsection{Preliminary}

At bidding time, the observable user-side features include the user profile $x_u$ and behavioral sequences, such as historical interactions $x_s^{(i)}$ and ad contexts $x_s^{(a)}$. 
In contrast, the onboarding content sequence is unavailable at bidding time.
We denote it by $x_c = \{v_0, v_1, \dots, v_n\}$, spanning up to $D$ days, where $v_i \in x_c$ corresponds to an item interacted with by user $u$ after conversion.

The prediction target $y$ is defined as the revisit frequency LT\_d of a user over the next $d$ days.
Therefore the retention prediction task at serving time can be formulated as the conditional probability $p(y | x_u)$, since $x_u$ is the only information available for online inference before conversions. 
However, during training, the sequence of onboarding content $x_c$ is accessible and can be leveraged as auxiliary information for retention modeling learning.





\subsection{Onboarding Content Leakage Encoding}

In the first stage, we aim to obtain high-quality onboarding content representations as supervision signals for subsequent distillation. To this end, we deliberately leak such information during training to learn the onboarding content encoder.

\textbf{Hierarchical Attention Encoder (HAE).}
Users may interact with items for multiple days after conversion, yielding long engagement sequences with hundreds of interactions. 
The raw onboarding content sequence $x_c$ may span multiple days with hundreds of interactions.
While such sequence reveals temporal evolution of user engagement, it also presents modeling and compression challenges. 
Notably, the sequence is inherently hierarchical, consisting of fine-grained interaction patterns within each day and higher-level behavioral evolution across days. 
To capture this structure, we adopt a two-level hierarchical attention encoder that models both intra-day dependencies and inter-day dynamics.

Specifically, we split the onboarding content sequence by calendar day $d \in \{1, \dots, D\}$, take the first $N$ interactions of each day and embed them into a sequence of item representations $\mathbf{H}^{(d)} = \{h^{(d)}_0, \dots, h^{(d)}_N\}$.
Then a cross-attention layer uses the user profile embedding $x_u$ as the query, with the item representations as keys and values:
\begin{equation}
    s^{(d)} = \text{CrossAttn}(x_u, \mathbf{H}^{(d)}, \mathbf{H}^{(d)}).
\end{equation}
Thus, we achieve the focused compression within the single day and obtain a compact semantic representation $s^{(d)}$.

The day-level representation sequence $\{s^{(1)}, \dots, s^{(D)}\}$ are then fed into a causal self-attention layer. By introducing a causal mask, each day is allowed to attend only to current and preceding representations, enabling the model to capture temporal dependencies across days without accessing future information:
\begin{equation}
    \tilde{s}^{(1)}, \dots, \tilde{s}^{(D)} = \text{CausalSelfAttn}(s^{(1)}, \dots, s^{(D)}).
\end{equation}
This causal structure prevents future leakage and preserves the chronological semantics of engagement evolution, since representations of earlier days are not influenced by subsequent interactions. 
Based on the resulting sequence, we use the representation of $d$ day $\tilde{s}^{(d)}$ for the LT\_d prediction task (\textit{e.g.}, $\tilde{s}^{(1)}$ for LT1 and $\tilde{s}^{(7)}$ for LT7), as it provides a compact summary of the user’s engagement history up to this day.

\textbf{Joint Training.} 
The onboarding content embedding after MLP projecting $e_c = \text{MLP}(\tilde{s})$ is concatenated with the original retention model's features, with the whole framework is jointly optimized under the retention prediction objective:
\begin{equation}
    \mathcal{L}_{Stage1} = \text{BCE}(f_{retention}([x_u; e_c]), y),
\end{equation}
where $f_{retention}(\cdot)$ is the retention model and $[;]$ denotes concatenation. Such end-to-end training encourages the learned onboarding content representation to better align with the retention task, ensuring the retention model's awareness of the onboarding content.

\subsection{User Representation Distillation Alignment}

In the second stage, we aim to train a user encoder, which maps observable user features to a user representation that approximates the onboarding content representation. 
Thus we froze the onboarding content encoder and align user representation through distillation.

\textbf{Sequence Fusion Encoder (SFE).}
Besides the user profile $x_u$, the observable user features also include multiple heterogeneous behavior sequences, such as historical interactions $x^{(i)}_s$ and ad contexts $x^{(a)}_s$. 
However, considering that the training sample is a re-engaged user, whose historical interactions are far removed from the bidding time, thus exhibiting severe temporal staleness.
Moreover, ad contexts are influenced by exposure and delivery mechanisms, which may carry limited personalized intent.
We therefore use the user profile as a unified anchor of user specific preference and concatenate with each item embedding $e^{(m)}_j$ of the heterogeneous sequence $x^{(m)}_s$, where $m \in \{i, a, \dots\}$.
Consequently, we have the user conditioned representation sequence:
\begin{equation}
    \mathbf{H}^{(m)} = \{h^{(m)}_0, \dots, h^{(m)}_n\}, ~ where ~ h^{(m)}_j = [e^{(m)}_j;x_u].
\end{equation}

We then use a small set of $K$ learnable query tokens together with sequence-specific Q-Former~\cite{blip} modules to compress multiple variable-length behavior sequences into fixed-length user interest representations:
\begin{equation}
    s^{(m)} = \text{Q-Former}(\mathbf{H}^{(m)}).
\end{equation}
In this way, heterogeneous sequences are mapped into a unified representation space. 
We then integrate the user-side features and construct task-specific towers for retention prediction over different horizons.
Each tower consisting of a feature interaction layer and a projection head to produce a user representation tailored to the retention task:
\begin{equation}
    e_u = \text{TaskTower}([x_u; s^{(m)}]).
\end{equation}


\textbf{Distillation Alignment.}
To achieve an approximate estimation of the invisible information, we adopt distillation to align user representation with onboarding content representation.
Concretely, for each task $t$, the output of the onboarding content encoder $e^{(k)}_c$ from the former stage serves as the teacher signal, while that of the user encoder $e^{(k)}_u$ serves as the student representation.
We then minimize the representation gap measured by a similarity-based loss $\mathcal{L}_{sim}$(\textit{e.g.}, cosine similarity loss):
\begin{equation}
    \mathcal{L}_{align} = \sum_{t \in \text{Task}} \mathcal{L}_{sim}(e^{(t)}_u, \text{sg}(e^{(t)}_c)),
\end{equation}
where $\text{sg}(\cdot)$ denotes the stop-gradient operator to stabilize the distillation target and prevent the collapse of the representation.

To ensure the representation alignment remains in service of the retention objective and preserves the stability of the core task, we further combine the aligned user representation with the retention model's features and jointly optimize:
\begin{equation}
    \mathcal{L}_{Stage2} = \text{BCE}(f_{retention}([x_u; e_u]), y) + \lambda \mathcal{L}_{align}.
\end{equation}

\subsection{Inference}

During inference, the framework relies only on the main retention model $f_{retention}(\cdot)$ and the latent representation of onboarding content  estimated by the user encoder $g_u(\cdot)$:
\begin{equation}
    \hat{y} = f_{retention}([x_u;g_u(x_u)]),
\end{equation}
while the onboarding content encoder is discarded after training. 
This design enables informative retention prediction without introducing feature leakage.

\section{Experiments}

\subsection{Experimental Settings}

To validate the effectiveness of our proposed framework, we conduct extensive offline and online experiments.

\textbf{Dataset.} 
We construct an industrial dataset collected from a widely used short-video platform to conduct offline analysis. The dataset includes millions of users and billions of interaction records.

\textbf{Evaluation.}
We define two retention prediction tasks, LT1 and LT7, corresponding to 1-day and 7-day engagement frequency respectively. 
We follow the widely used evaluation metrics, AUC and GAUC, to evaluate model performance.

\textbf{Implementation.}
We ensure that all features are visible during training, but onboarding content is unavailable during evaluation.
We adopt PPNET~\cite{ppnet} as the framework of the retention model. 

\subsection{Offline Experiments Analysis}
\label{subsec:offline}

\begin{table}[t!]
  \caption{Overall performance of the model with different training stage. The performance of the full model is bold, and the upper bound is \underline{underline}.}
  \vspace{-0.3cm}
  \label{tab:offline}
  \begin{tabular}{lcccc}
    \toprule
    \multirow{2.5}{*}{\textbf{Method}} & \multicolumn{2}{c}{\textbf{LT1}} & \multicolumn{2}{c}{\textbf{LT7}} \\
    \cmidrule(r){2-3} \cmidrule(r){4-5} & AUC & GAUC & AUC & GAUC \\
    \midrule
    \textbf{Base} & 0.7297 & 0.7227 & 0.6903 & 0.6909 \\
    \midrule
    \textbf{OCARM} & & & & \\
    \quad \textbf{\textit{w}/} Stage 1 (Upper Bound) & \underline{0.7468} & \underline{0.7371} & \underline{0.7002} & \underline{0.7007} \\
    \quad \textbf{\textit{w}/} Stage 2  & 0.6709 & 0.6719 & 0.6430 & 0.6435 \\
    \quad \textbf{\textit{w}/} Stage 1 + 2 (Full) & \textbf{0.7369} & \textbf{0.7311} & \textbf{0.6949} & \textbf{0.6957} \\
    \bottomrule
  \end{tabular}
\end{table}

\begin{table}[t!]
  \caption{Performance gains over the base model of several variants with different encoder architecture.}
  \vspace{-0.3cm}
  \label{tab:ablation}
  \begin{tabular}{ccccc}
    \toprule
    \multirow{2.5}{*}{\textbf{Variant}} & \multirow{2.5}{*}{Content Encoder} & \multirow{2.5}{*}{User Encoder} & \multicolumn{2}{c}{\textbf{$\Delta$AUC (\%)}} \\
    \cmidrule(r){4-5} & & & LT1 & LT7 \\
    \midrule
    \textbf{Base} & - & - & -  & - \\
    \midrule
    \textbf{Variant 1} & MLP & MLP & +0.35 & +0.23 \\
    \textbf{Variant 2} & HAE & MLP & +0.56  & +0.32 \\
    \textbf{Variant 3} & HAE & SFE & +0.72  & +0.46 \\
    \bottomrule
  \end{tabular}
  \vspace{-0.3cm}
\end{table}


\begin{figure}[t!]
    \centering
    \subfloat[LT1]{
        \includegraphics[width=4.2cm]{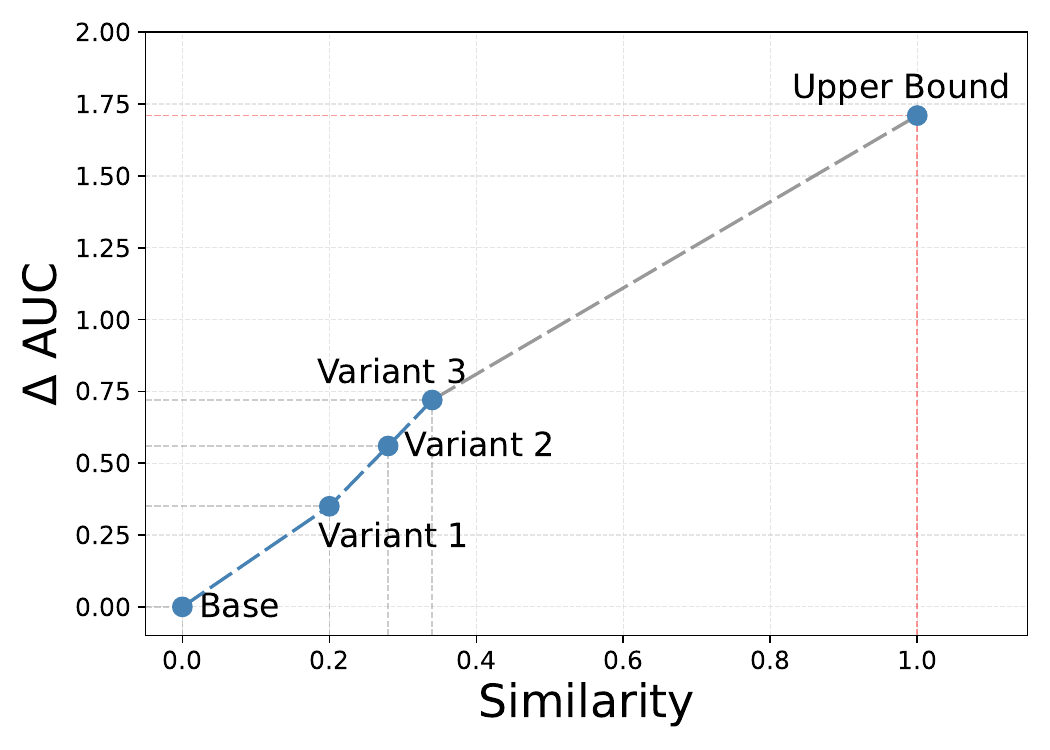}
    }
    \subfloat[LT7]{
        \includegraphics[width=4.2cm]{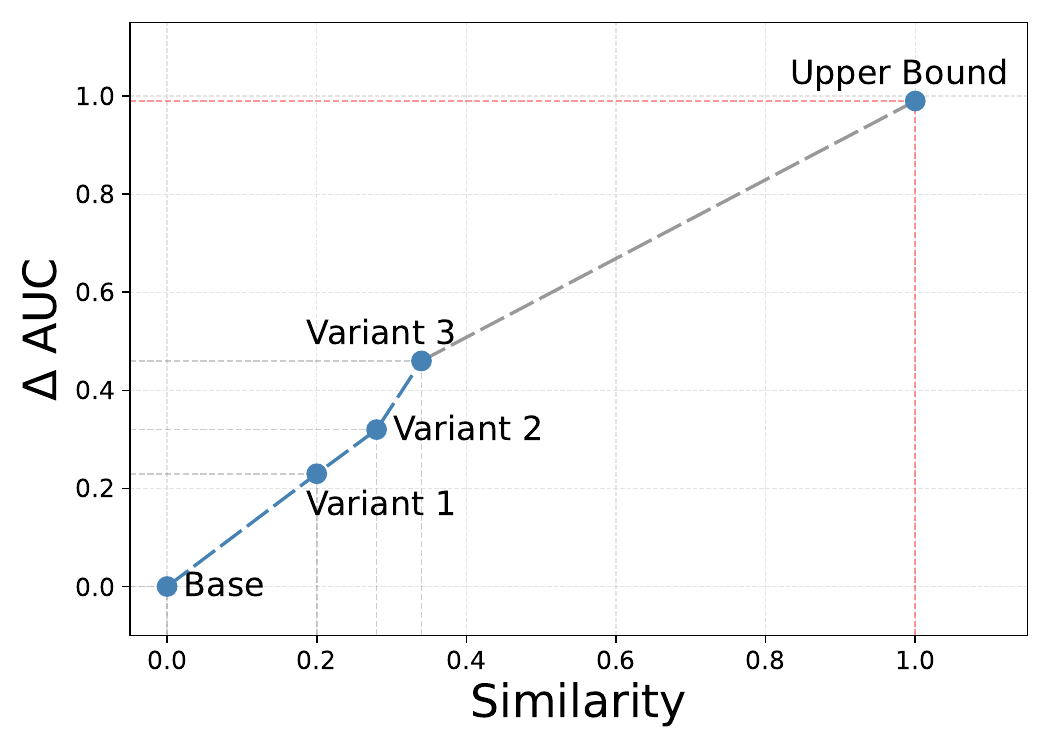}
    }
    \vspace{-0.2cm}
    \caption{Performance gains of retention task from increased representation similarity.}
    \vspace{-0.3cm}
    \label{fig:improve}
\end{figure}

Table \ref{tab:offline} summarizes the offline results with different training-stage configurations. 
Using Stage 1 alone, where onboarding content is deliberately leaked during training, yields substantial gains over the base model. 
This confirms that onboarding content provides strong informative signals for retention prediction and can be viewed as an upper bound.
Using Stage 2 alone by jointly training HAE (without stop-gradient) and SFE leads to significant performance degradation.
A possible reason is that, without a well-learned teacher representation, direct alignment introduces unstable optimization and even representation collapse, thereby degrading alignment learning.
In comparison, the full two-stage framework first learns reliable onboarding content representations and then transfers them through distillation, enabling the model to approximate inaccessible information at inference time. 
As a result, it achieves clear improvements over the base model while avoiding feature leakage, validating the effectiveness of our two-stage distillation-aligned framework.

We further investigate the performance of several variants with different encoder architectures:
\begin{itemize}
    \item Variant 1 adopts MLP for content encoder and user encoder.
    \item Variant 2 replaces the MLP of content encoder with HAE.
    \item Variant 3 (Full) further replaces the user encoder with SFE.
\end{itemize}
Table \ref{tab:ablation} shows that stronger encoder architectures consistently improve retention prediction, suggesting that better network designs more effectively exploit onboarding content signals. 
Figure \ref{fig:improve} further provides mechanistic evidence for this trend.
As the similarity between the user representation and the onboarding content representation increases, the performance grows monotonically. 
This suggests that variants with better alignment achieve larger retention gains.
Meanwhile, the remaining gap to the upper bound indicates that approximating inaccessible information still leaves substantial room for improvement and exploration.



\subsection{Online A/B Test}

\begin{table}[t!]
  \caption{Online A/B test result of real industrial scenario.}
  \vspace{-0.3cm}
  \label{tab:ab}
  \begin{tabular}{ccc}
    \toprule
    \textbf{Metric} & \textbf{Non-Uninstalled} & \textbf{Uninstalled} \\
    \midrule
    \textbf{Re-engaged Devices} & +20.468\% & +34.430\% \\
    \midrule
    \textbf{LT30} & +11.548\% & +22.179\% \\
    \bottomrule
  \end{tabular}
  \vspace{-0.5cm}
\end{table}

We deploy our framework in the RTB system of a short-video platform, with the result during multiple days of A/B test shown in Table \ref{tab:ab}.
The result shows that our method consistently improves retention prediction across diverse user groups. 
Notably, the gains are much larger for uninstalled users than for non-uninstalled users, suggesting that our approach is particularly effective at identifying the long-term value of hard-to-retain users. 
Overall, these findings demonstrate the effectiveness and practical value of our method in real industrial scenario.

\section{Conclusion}

In this paper, we identify the temporal visibility gap in RTB retention prediction, where the most predictive onboarding content is inaccessible at bidding time. We then propose OCARM, a two-stage distillation-aligned framework that first encodes leaked onboarding content via hierarchical attention as teacher signals, then trains a user encoder to approximate these representations from observable features alone. Offline and online experiments on a real-world short-video platform validate the effectiveness of our framework.

\balance
\bibliographystyle{ACM-Reference-Format}
\bibliography{sample-base-extend.bib}

\end{document}